\documentclass[12pt]{article} \usepackage{amsmath}
\usepackage{amssymb} \usepackage{amsthm} \usepackage{verbatim}
\usepackage{graphicx}\usepackage{hyperref}

\title{Non-local Boxes} \author{Philippe Lamontagne\\[1em] Université de Montréal} \date{This report was prepared in 2011\\ as part of a research internship}

\begin{document}
\maketitle
\pagebreak

\tableofcontents
\pagebreak

\newtheorem{thm}{Theorem} \newtheorem{defn}{Definition}
\newtheorem{cor}{Corollary} \newtheorem{lem}{Lemma}

\section{Introduction}

The study of non-local boxes arose from the study of quantum
entanglement and from the question: ``why isn't entanglement more
non-local?''. Correlations stronger than quantum entanglement, but that
still do not allow for instantaneous transmission of information have
been known to exist \cite{springerlink:10.1007/BF02058098}.

\subsection{Preliminaries}

The concept of non-local boxes is inspired by that of quantum
systems. They are closely related as a quantum system can be viewed as
a non-local box, where the choice of measurement is the input and the
outcome of the measurement is the output, and a non-local box can be
viewed as a super-quantum system. Of course, not all boxes as defined
under are non-local: they can be local, quantum, or super-quantum.

\begin{defn}\label{defbox}
A bipartite correlated box (or box) is a device with two ends, one of
which is held by Alice, the other one by Bob.  Each end has the
following input-output behaviour: given input $x$ on Alice's side
(respectively $y$ on Bob's side), the box will output $a$
(respectively $b$) according to some probability distribution
$P(a,b|x,y)$ where $x,y,a,b \in \{0,1\}$.
\end{defn}

Throughout this paper, we will refer to boxes by their probability
distributions. For convenience, we will also write $P(ab|xy)$
and $P(a,b|x,y)$ interchangeably.

It is important to note that boxes are atemporal, meaning that the
output comes out on one side as soon as an input is given. Was this
not the case (if, for example, the box waits for both inputs before
giving outputs), then one could transmit information to the other
party by deliberately delaying it's input.

Quantum entanglement does not allow for faster-than-light
communication. This property is called non-signalling. Likewise, we are
only interested in studying boxes that are non-signalling, which means
Alice cannot learn anything from Bob's input by looking at her output.

\begin{defn}\label{non-signalling}
A box $P$ is \emph{non-signalling} if the sum over Bob's inputs of the
joint probability distribution is equal to Alice's marginal
distribution and vice versa:
\begin{align*}
&\sum_b P(a,b|x,y) = \sum_b P(a,b|x,y')= P^A(a|x) \quad\forall
  a,x,y,y' \text{ and}\\&\sum_a P(a,b|x,y) = \sum_a P(a,b|x',y)=
  P^B(b|y) \quad\forall b,x,x',y.
\end{align*}
It is \emph{signalling} if it is not non-signalling.
\end{defn}

Non-signalling correlations can be of many types. Of the boxes with
this property, we find ones that can be implemented with classical
theory, quantum theory or even super-quantum theory. Since the class
of non-signalling correlations include quantum and classical ones, we
may define a box as being non-local in the same way some quantum
correlations are non-local. A box is said local if the output on one
side depends only on the input on the same side. Local correlations
can be simulated with only shared randomness by non-communicating
participants.

\begin{defn}\label{f9qh4233dga}
A box $P$ is \emph{local} if it can be written as
\[
P(a,b|x,y)=\sum_i \lambda_i P^A_i(a|x)P^B_i(b|y)
\]
where $\lambda_i\geq 0$ and $\sum_i \lambda_i=1$.  A box is
\emph{non-local} if it is not local.
\end{defn}

In essence, definition \ref{f9qh4233dga} says that any local box is a
convex combination of local boxes. This is in accordance with the fact
that the set of local correlations form a
polytope\cite{2005PhRvA..71b2101B} with the vertices being
deterministic boxes (i.e. boxes with output uniquely determined).

Now that we have defined what is non-locality, it would be useful to
be able to quantify it. The value defined next is taken from the
Clauser-Horne-Shimony-Holt inequality (or CHSH
inequality)\cite{PhysRevLett.23.880} which give an upper bound on
local correlations. This inequality was designed as an application of
Bell's famous theorem\cite{bell_js:1964a}, but became a measure of
non-locality. It was originally stated with expectation values of
measurements of quantum system. We give a more information theoretical
description from \cite{PhysRevLett.102.120401}.

\begin{defn}\label{CHSH}
  Let $X_{xy}(P)= P(00|xy)+P(11|xy)-P(01|xy)-P(10|xy)$.  The CHSH
  value of box $P$ is
  \[
    CHSH(P) = \max_{xy} |X_{xy}(P)+X_{x\bar y}(P)+X_{\bar x
      y}(P)-X_{\bar x\bar y}(P)|
  \]
\end{defn}

Clause, Horne, Shimony and Holt's derivation of Bell's theorem is stated
in theorem \ref{l12jk.o8efhasp90qawuh3f} as a upper bound on the
correlation of two local variables. It gives a necessary and
sufficient condition on correlations for them to be local.

\begin{thm}[Bell] \label{l12jk.o8efhasp90qawuh3f}
  A box $P$ is local if and only if $CHSH(P)\leq 2$.
\end{thm}

Cirel'son\cite{springerlink:10.1007/BF00417500} later found an upper
bound on the CHSH value that all quantum correlations must obey. It is
a necessary condition for correlations to be achievable by quantum
mechanics. 

\begin{thm}[Cirel'son]
  If a box $P$ can be implemented by quantum mechanics, then
  $CHSH(P)\leq 2\sqrt2$.
\end{thm}

However, this condition is not sufficient. This was remedied by
\cite{springerlink:10.1007/BF00732549} who found a necessary and
sufficient condition on boxes for them to be quantum.

\begin{thm}
  A box $P$ can be implemented by a quantum state if and only if
  $|\arcsin X_{xy} + \arcsin X_{x\bar y} + \arcsin X_{\bar x y} -
  \arcsin X_{\bar x \bar y}| \leq \pi$. For any $xy= 00, 01, 10, 11$
  where $X_{xy}$ is defined in definition \ref{CHSH}.
\end{thm}

The following box was introduced by Popescu and Rohrlich
\cite{springerlink:10.1007/BF02058098} as a correlation achieving the
maximal algebraic of 4 of the CHSH inequality. It is at the core of
the study of non-locality and is used in the proofs of many of the
results presented in this work.

\begin{defn}
The Popescu-Rohrlich box (PR-box) is described by the following
probability distribution:
\[
P^{PR}(a,b|x,y)=
\begin{cases}
1/2 &\text{ if } a\oplus b=xy \\ 0 &\text{ otherwise.}
\end{cases}
\]
The noisy symmetric (or isometric) PR-boxes are the boxes of the form
\[
P_\epsilon = \epsilon P^{PR} + (1-\epsilon)P^{\overline{PR}}.
\]
where $P^{\overline{PR}}$ is the anti-PR-box: $P^{\overline{PR}}(ab|xy)
= 1/2$ if $a\oplus b \neq xy$.
\end{defn}

\section{The Non-Signalling Polytope}\label{nspoly}

Barrett et al.\cite{2005PhRvA..71b2101B} characterized the class of
non-signalling correlations.  All probability distributions within this
class are subject to the following conditions:
\begin{enumerate}
\item positivity
\[
P(ab|xy)\geq 0;
\]
\item normalization
\[
\sum_{a,b}P(ab|xy)=1;
\]
\item non-signalling constraints (see definition \ref{non-signalling}).
\end{enumerate}
Since these constraints are linear, the class forms a polytope. To
determine the dimension of the polytope, first note that the set of
probabilities $P(a,b|x,y)$ where $x,y,a,b\in \{0,1\}$ form a table
  with $2^4$ entries. The dimension of the polytope is then given by
  subtracting the number of independent constraints from $2^4$ which
  gives us the number of in-dependant ``variables'' of the table, and
  turns out to be 8.

The polytope has 24 vertices, 16 of which correspond to local
deterministic boxes of the form
\[
P^{\alpha\beta\gamma\delta}(a,b|x,y)=\begin{cases} 1 &\text{ if } a=
\alpha x \oplus \beta, \;b=\gamma y\oplus \delta\\ 0 &\text{otherwise}
\end{cases}
\]
where $\alpha,\beta,\gamma,\delta\in \{0,1\}$.  These alone form the
  local polytope containing all local boxes as a convex combination of
  those 16 vertices.  The remaining 8 vertices of the non-local
  polytope are of the form
\[
P^{\alpha\beta\gamma}(a,b|x,y)=\begin{cases} 1/2 &\text{ if } a\oplus
b=xy\oplus \alpha x\oplus \beta y\oplus \gamma \\ 0 &\text{otherwise}
\end{cases}
\]
where $\alpha,\beta,\gamma\in \{0,1\}$.

\begin{thm}
All vertices of the local polytope are equivalent under reversible
local operations and all non-local vertices of the non-signalling
polytope are equivalent under reversible local operations.
\end{thm}

By \emph{reversible local operations}, it is meant that Alice may
relabel her input, $x\leftarrow x\oplus 1$, or she may relabel her
output conditionally on her input, $a\leftarrow a\oplus \alpha x\oplus
\beta$, and similarly for Bob. It is easy to see that any vertex of a
given class (local or non-local) can be transformed into any other
vertex of the same class by these operations.

\subsection{Depolarization}\label{sec:depolarization}

Demoralization is the act of taking a box which is a mixture of any
non-signalling box and transforming it into a symmetric box, while
preserving the CHSH value.

It consists of generating three maximally random bits
$\alpha,\beta,\gamma$ and doing the following substitutions: $x
\rightarrow x \oplus \alpha$, $y \rightarrow y \oplus \beta$, $a
\rightarrow a\oplus \beta x \oplus \alpha\beta \oplus \gamma$ and $b
\rightarrow b\oplus \alpha y \oplus \gamma$.

Note that this operation requires three bits of shared randomness
between the two parties for every box they wish to depolarize.

\section{Trivial Communication Complexity}\label{seccomcompl}

There has been evidence that non-locality helps in the communication
complexity of some distributed tasks. See for example
\cite{springerlink:10.1007/s10701-005-7353-4}. Protocols that make use
of non-locality in the form of quantum entanglement offers advantages
over local protocols, but since quantum non-locality is restricted, it
is natural to ask ourselves if stronger non-locality is more helpful.

\begin{defn}
The communication complexity of a function $f$ is \emph{trivial} if it
can be computed using a single bit of communication per participant.
\end{defn}

This is the minimum communication needed to compute any function which
is not itself trivial (i.e. it does not depend on only one of the
inputs). 

\subsection{Two participants}

It will be useful to define the following property. Most proofs of
trivial communication complexity using non-local boxes try to achieve
this property.

\begin{defn}
The Boolean function $f$ is distributively computed by Alice and Bob
if they respectively receive $x$ and $y$ and output $a$ and $b$ such
that $a\oplus b=f(x,y)$.
\end{defn}
Evidently, every function that can be distributively computed has
trivial communication complexity.  So every function that has
communication complexity strictly greater than 1 cannot be
distributively computed, and since most functions have non-trivial
communication complexity, most functions are not distributively
computable.  Perhaps surprisingly, the next result by van Dam
\cite{2005quant.ph..1159V} shows that the existence of the NLB renders
every function's communication complexity trivial.
\begin{thm}
  In a world in which perfect non-local boxes exists, all Boolean
  functions can be distributively computed.
\end{thm}

The proof uses the fact that every function $f:\{0,1\}^n\times
\{0,1\}^n \to \{0,1\}$ can be expressed as a multivariate polynomial
which can be written in the form $f(x,y) = \sum_i P_i(x) \cdot
Q_i(y)$, where $P_i$ and $Q_i$ are polynomials and $x,y \in
\{0,1\}^n$. This can be distributively computed by Alice and Bob
because $P_i(x)$ depends only of $x$ and $Q_i(y)$ only of $y$. They
then input $P_i(x)$ and $Q_i(y)$ into the $i$th box.

\subsection{$n$ participants}
Let us extend the definition of distributed computation to $n$
players, where the parity of the outputs is equal to the value of the
function. 
\begin{defn}
The Boolean function $f$ is $n$-partite distributively computed by $n$
participants if they respectively receive $x_i$ and output $a_i$,
$1\leq i\leq n$, such that $\bigoplus_{i=1}^n a_i=f(x_1,\ldots,x_n)$.
\end{defn}

The next result, by Barrett and Pironio \cite{PhysRevLett.95.140401},
extends van Dam's result to $n$-partite communication complexity.

\begin{thm}
Correlations of the form
\[
P(a_1,\ldots,a_n|x_1,\ldots,x_n) =
\begin{cases}
  1/2^{n-1} &\text{if } \bigoplus_{i=1}^n a_i=f(x_1,\ldots,x_n) \mod 2
  \\ 0 & \text{otherwise}
\end{cases}
\]
can be simulated with non-local boxes.
\end{thm}

\begin{cor}
Any $n$-partite communication complexity problem can be solved with
$n-1$ bits of communication.
\end{cor}
This is easy to see, as all participants send their outputs to the
first who can then compute the function.

\subsection{Probabilistic}

All triviality results presented thus far concern deterministic
multipartite functions. Brassard et al. \cite{PhysRevLett.96.250401}
found that this still applies when considering probabilistic
multipartite computation.

\begin{thm}
In a world in which noisy non-local boxes which succeed more than
$\frac{3+\sqrt 6}6 \approx 90.8\%$ exist, all probabilistic functions
can be distributively computed.
\end{thm}

This lets us define the set of boxes that trivialize communication
complexity.

\begin{cor}\label{Bcc}
  Let $B_{cc}=4\sqrt{2/3}\approx 3.266$, then all boxes $P$ such that
  $CHSH(P)>B_{cc}$ trivialize communication complexity.
\end{cor}

The CHSH value of a symmetric non-local box with probability of
success $\frac{3+\sqrt6}{6}$ is $4\sqrt{2/3}$ and using the
depolarization protocol described in section \ref{sec:depolarization},
all boxes above CHSH value $B_{cc}$ trivialize communication
complexity.

\section{Non-local Games} 

All the results of this section are due to Cleve et
al.\cite{2004quant.ph..4076C}, except for the ones of section
\ref{secnonlocalcomp}. Those last are from Linden et
al. \cite{PhysRevLett.99.180502}. 

When playing a non-local game, Alice and Bob are space-like separated
but allowed to share randomness. They are, however, allowed to elaborate
a strategy beforehand. Alice and Bob respectively receive $x\in X$ and
$y\in Y$ picked at random according to the probability distribution
$\pi$. They must respectively output $a\in A$ and $b\in B$. They win
if $V(a,b,x,y)=1$.

\begin{defn}
A non-local game $G=(X\times Y,A\times B,\pi,V)$ consists of a set of
inputs $X\times Y$, a set of outputs $A\times B$, a probability
distribution $\pi:X\times Y\to[0,1]$ and a predicate $V:X\times
Y\times A\times B\to \{0,1\}$.
\end{defn}

Next is defined the best probability with which Alice and Bob can win
a game when they are restricted to classical strategies, i.e.,
strategies that do not make use of non-locality.

\begin{defn}
The maximum winning probability for a classical strategy for a
non-local game $G=(X\times Y,A\times B,\pi,V)$ is
\[
\omega_C(G) = \max_{a,b} \sum_{x,y} \pi(x,y) V(a(x),b(y),x,y)
\]
where the maximum is taken over all functions $a:X\to A$ and $b:Y\to
B$.
\end{defn}

If we allow Alice and Bob to share entanglement, their winning
probability may benefit from it. A quantum strategy is determined by a
bipartite state $|\psi\rangle$ shared between Alice and Bob. They both
perform some measurement according to their respective input and
output the result of that measurement.

precisely, a quantum strategy consists of:
\begin{itemize}
\item a state $|\psi\rangle \in \mathcal{A}\otimes \mathcal{B}$ for
  $\mathcal{A}$ and $\mathcal{B}$ isomorphic copies of the vector
  space $\mathbb{C}^n$ for some $n$. Where $\mathcal{A}$ represents
  Alice's part of $|\psi\rangle$ and $\mathcal{B}$ Bob's part;
\item two sets of positive semidefinite $n\times n$ matrices
\[
\{X_x^a| x\in X,a \in A\}\text{ and } \{Y_y^b|y\in Y,b\in B\}
\]
satisfying
\[
\sum_{a\in A}X_x^a = \mathbb{I} \text{ and } \sum_{b\in
  B}y_y^b=\mathbb{I}
\]
for every $x\in X$ and $y\in Y$, where $\mathbb{I}$ is the $n\times n$
identity matrix.
\end{itemize}

We define the maximum winning probability of players with quantum
strategies the following way.

\begin{defn}
  The maximum winning probability of a quantum strategy for a
  non-local game $G=(X\times Y,A\times B,\pi,V)$ is
\[
\omega_Q(G) = \max_{|\psi\rangle} \sum_{(x,y)\in X\times Y}\pi(x,y)
\sum_{(a,b)\in A\times B}\langle\psi| X_x^a\otimes Y_y^b|\psi\rangle
V(a,b,x,y)
\]
\end{defn}

\subsection{Binary Games}

In this section, we consider non-local games where answers are bits. 

\begin{defn}
  A binary game $G=(X\times Y,A\times B,\pi,V)$ is a non-local game
  where $A=B=\{0,1\}$.
\end{defn}

This next result states that quantum strategies cannot have an
advantage over classical strategies if there exists a quantum strategy
that always win the game.

\begin{thm}
  Let $G$ be a binary game. If there exists a quantum strategy for $G$
  that wins with probability 1, then $\omega_C(G)=1$.
\end{thm}

This result is fairly strong, it implies that we will never be able to
perfectly achieve such tasks when it is not possible classically.

\subsection{XOR-games}	

In this section, we study games for which the result depends not on
the individual answers, but on the exclusive-OR of respective
answers. This category of games include the bipartite communication
complexity tasks of section \ref{seccomcompl}.

\begin{defn}
  A XOR-game $G=(X\times Y,A\times B,\pi,V)$ is a binary game where
  $V:C\times X\times Y\to \{0,1\}$ and $C=\{a\oplus b|a\in A,b\in
  B\}$.
\end{defn}

The following definition will be of use for some of the results of
this section. It is the winning probability when players are
restricted to a trivial strategy, a trivial strategy consisting of
outputting random bits.  

\begin{defn}
  The success probability for a game $G$ if both parties are
  restricted to a trivial strategy (output random bits) is
\[
\tau(G)=\frac12 \sum_{c\in\{0,1\}}\sum_{x,y}\pi(x,y)V(c,x,y)
\]
\end{defn}

When playing a XOR-game, the gain of the best quantum strategy over
the trivial strategy cannot be too great compared to the gain of the
best classical strategy over the trivial strategy. This is the essence
of the following result, which upper bounds the gap between quantum and
classical advantages over the trivial strategy.

\begin{thm}
  Let $G$ be a XOR-game. Then
\[
\frac{\omega_Q(G)-\tau(G)}{\omega_C(G)-\tau(G)}\leq K_G
\]
where $K_G$ is Grothendeick's constant.
\end{thm}

Grothendeick's constant $K_G$ is the smallest number such that, for
all integers $N\geq 2$ and all $N\times N$ real matrices $M$, if
\[
\left|\sum_{i,j}M(i,j)a_ib_j\right| \leq 1
\]
for all numbers $a_1,\ldots,a_N,b_1,\ldots,b_N$ in $[-1,1]$, then
\[
\left|\sum_{i,j}M(i,j)\langle u_i|v_j\rangle \right| \leq K_G
\]
for all unit vectors $|u_1\rangle, \ldots, |u_N\rangle, |v_1\rangle,
\ldots, |v_N\rangle$ in $\mathbb{R}^n$ for any $n$.

The exact value of Grothendieck's constant is not known, but it is
known to satisfy
\[
1.6769\leq K_G \leq \frac\pi{2\log(1+\sqrt2)}\approx 1.7822.
\]

Finally, the coming result upper bounds the maximum quantum winning
probability by a function of the maximum classical winning probability.

\begin{thm}
  Let $G$ be a XOR-game. Then
\[
\omega_Q(G)\leq \begin{cases} \gamma_1\omega_C(G) &\text{if
  }\omega_C(G)\leq \gamma_2 \\ \sin^2(\frac\pi2 \omega_C(G)) &\text{if
  }\omega_C(G)>\gamma_2,
\end{cases}
\]
where $\gamma_1$ and $\gamma_2$ are the solution to the equation
$\frac \pi 2 \sin (\pi \gamma_2) = \frac{\sin^2(\frac \pi 2 \gamma_2)}
{\gamma_2} = \gamma_1$. $\gamma_1 \approx 1.1382$ and $\gamma_2
\approx 0.74202$.
\end{thm}

\subsection{Non-local Computation}\label{secnonlocalcomp}

Consider the scenario in which Alice and Bob wish to distributively
compute a function whose input is also distributed. Alice and Bob
respectively receive bit strings $x$ and $y$ and they must output
single bits $a$ and $b$ such that $a\oplus b=f(x\oplus y)$. What is
particular in this type of non-local game is that neither of the
players learn anything about the input since the individual bits of
$x$ and $y$ are uniformly distributed from Alice and Bob's
perspective. 

\begin{defn}
  A non-local computation game (or NLC-game) of a function $f$ is a
  XOR-game $G=(X\times Y,A\times B,\pi,V)$ where $V:C\times Z \to
  \{0,1\}$, $Z=\{x\oplus y|x\in X, y\in Y\}$ and $V(a\oplus b,x\oplus
  y)=1$ if $a\oplus b=f(x\oplus y)$.
\end{defn}

Linden et al. showed that when considering such a model, neither
classical nor quantum strategies can always win a given game.

\begin{thm}
  Let $G$ be a non-local computation game. Then
  \[
  \omega_C(G)=\omega_Q(G)<1
  \]
\end{thm}

\section{Non-locality Distillation}

The motivation behind the study of non-locality distillation is the
question of whether we can use a set of boxes to simulate the
behaviour of a more non-local one. For example, can we use a set of
$n$ noisy PR-boxes to simulate the behaviour of a less noisy PR-box. 

\begin{defn}
A non-locality distillation protocol (NDP) consists of local
operations performed by Alice and Bob on their respective ends of
$n$ boxes with a given CHSH value to simulate the input-output
behaviour of a higher valued box.  A non-locality distillation
protocol $\mathcal{N}$ on $n$ boxes $P$, denoted $\mathcal{N}^n[P]$, consists of
local operations performed on the boxes to simulate the
input/output behaviour of a box $P'=\mathcal{N}^n[P]$.
\end{defn}

Of course, for a distillation protocol to be useful we must have that
the CHSH value of the box simulated by the protocol is greater than
the CHSH value of the input boxes (i.e. $CHSH(P')>CHSH(P)$). However,
we do not require that distillation is achieved for all families of
boxes, because as we will soon see, this would be impossible.

\subsection{Limits on Distillation Protocols}

This section contains upper bounds and impossibilities on non-locality
distillation protocol. For example, the first statement of theorem
\ref{limits} asserts that no distillation protocol can create
non-locality from locality. The results of this theorem were taken
from \cite{2008arXiv0808.3317D}.

\begin{thm}\label{limits}
For any non-locality distillation protocol $\mathcal{N}$,
\begin{itemize}
\item if $CHSH(P)\leq 2$ then $CHSH(\mathcal{N}[P])\leq 2$;
\item if $P$ is a box whose correlations are achievable by quantum
  mechanics, then $CHSH(\mathcal{N}[P]) \leq 2\sqrt2$;
\item if $CHSH(P)<4$ then $CHSH(\mathcal{N}[P])<4$.
\end{itemize}
\end{thm}

It is important to understand that the second statement of theorem
\ref{limits} applies only to correlations that can be obtained by
measurements on quantum states. As we will see in section
\ref{sec:FWW}, some protocols bring boxes of CHSH value near $2$ and
brings them to $3 > 2\sqrt 2$, but these cannot be simulated by
measurements on quantum states.

Short\cite{PhysRevLett.102.180502} proved the impossibility of
distillation protocols operating on two copies of noisy PR-boxes.

\begin{thm}
  Two copies of a noisy PR-box cannot be distilled. For any
  $P_\epsilon = \epsilon P^{PR} + (1-\epsilon)P^{\overline{PR}}$,
  there is no $\mathcal{N}^2$ such that
  $CHSH(\mathcal{N}^2[P_\epsilon]) > CHSH(P_\epsilon)$.
\end{thm}

His proof, which applies to more general frameworks than just
non-local boxes, works by showing that the probability that the
protocol simulates a PR-box as a function of the same probability for
the initial boxes is a polynomial of degree two in the original
probability. He then shows a set of constraints that no polynomial of
degree two can satisfy.

\subsection{Known Distillation Protocols}

In this section, we consider only protocols achieving distillation of
correlations outside the quantum set.

All known distillation protocols are applied to the same family of
boxes, termed \emph{correlated non-local boxes} by Brunner and
Skrzypczyk\cite{PhysRevLett.102.160403}. Correlated non-local boxes
are of the form $P^C_\epsilon = \epsilon P^{PR} + (1-\epsilon) P^C$
where $P^C$ is the fully correlated box $P^C(ab|xy)=1/2$ if $a\oplus
b=0$. Correlated non-local boxes have a CHSH value of
$2(\epsilon+1)>2$. What characterizes these boxes is their bias
towards correlated outputs, i.e. $a\oplus b=0$. This means that when
the box outputs uncorrelated bits, you are assured that it has output
the correct answer. Both protocols presented here will make use of
this fact.

Non-locality distillation protocols can however be applied to any box
of the non-signalling polytope. Whether a protocol distills or not a
given box depends on its joint probability distribution.

\subsubsection{Forster, Winkler, Wolf Protocol}\label{sec:FWW}

The first non-locality distillation protocol for non-local boxes was
discovered by Forster, Winkler and Wolf
(FWW)\cite{PhysRevLett.102.120401}. Their protocol is fairly simple,
it uses the parity of the output of the initial boxes as output.

$\mathcal{F}^n[P^C_\epsilon](x,y)$
\begin{enumerate}
\item On inputs $x$ and $y$, input $x$ and $y$ into all $n$ boxes;
\item Let $a_i$ and $b_i$ be the outputs of the $i$th box, output
  $a=\bigoplus_{i=1}^n a_i$ and $b=\bigoplus_{i=1}^n b_i$.
\end{enumerate}

This protocol achieves distillation.

\begin{thm}
  For $n>1$ and $0<\epsilon<1/2$, $CHSH(\mathcal{F}^n[P^C_\epsilon]) =
  3-(1-2\epsilon)^n > 3-(1-2\epsilon)= CHSH(P^C_\epsilon)$.
\end{thm}

Perhaps interestingly, Peter Hoyer and Jibran Rashid showed in
unreleased work that when restricted to input $x$ and $y$ into all
boxes, the FWW protocol is optimal.

\subsubsection{Brunner, Skrzypczyk Protocol}
This protocol, introduced in \cite{PhysRevLett.102.160403}, operates
on two boxes. Unlike the FWW protocol which brings correlated value to
a CHSH value of 3 in the asymptotic limit, the Brunner Skrzypczyk
protocol brings then to the CHSH value of 4 in the asymptotic
limit. Which means they cross the communication complexity bound
$B_{cc}$, increasing the class of correlations that trivialize
communication complexity.

$\mathcal{B}^2[P^C_\epsilon](x,y)$
\begin{enumerate}
\item Input $x,y$ into first box;
\item Let $a_1$ and $b_1$ be the outputs of the first box, input
  $x\cdot a_1$ and $y \cdot b_1$ into second box;
\item Let $a_2$ and $b_2$ be the outputs of the second box, output $a=
  a_1\oplus a_2$ and $b= b_1 \oplus b_2$.
\end{enumerate}

\begin{thm}
  For $0<\epsilon<1$, $CHSH(\mathcal{B}^2[P_\epsilon^C])=3\epsilon -
  \epsilon^2 + 2 > 2(\epsilon + 1) = CHSH(P^C_\epsilon)$.
\end{thm}

When applied to boxes of the form $\epsilon P^{PR}+ \delta
P^{\overline{PR}} + (1- \epsilon - \delta)P^C$ for
$0<\delta<\epsilon<1$, which are achievable by quantum states, the
protocol still achieves distillation for some values of $\epsilon$ and
$\delta$ (without crossing tsirelson's bound of course).

\begin{cor}
  There exists correlations arbitrarily close to the classical and
  quantum sets of correlations that trivialize communication
  complexity.
\end{cor}

The Brunner Skrzypczyk protocol brings boxes of CHSH value arbitrarily
close to 2, yet still unreachable by quantum states, and distills then
to CHSH value arbitrarily close to 4 crossing the bound $B_{cc}$
defined in corollary \ref{Bcc}


\section{Implications in Cryptography}
\begin{defn}
  An oblivious transfer (OT) protocol is a protocol in which a sender
  sends a message to the receiver with probability $1/2$, while
  himself learning nothing of whether the receiver received the
  message. One out of two oblivious transfer (1-2 OT) is a variant in
  which the sender holds two bits $s_0$ and $s_1$, and the receiver
  has bit $c$. The receiver wishes to learn bit $s_c$ without the
  sender learning $c$.
\end{defn}

Wolf and Wullschleger \cite{2005quant.ph..2030W} gave a protocol for
secure 1-2 OT. Their protocol uses a single PR-box and proceeds as
follows. Alice inputs $x=x_0\oplus x_1$. Bob inputs $y=c$. Alice gets
output $a$ and Bob $b$. Alice sends $m=x_0\oplus a$ to Bob. Bob
computes $m\oplus b = x_0\oplus a\oplus b= x_0\oplus (x_0\oplus x_1)
c=x_c$.

Wolf and Wullschleger's protocol for 1-2 OT is secure, but when trying
the usual reduction from OT to 1-2 OT, it becomes insecure. In the
reduction, the sender uses $s_k=b$ and $s_{\bar k}=0$ with
$k\in_R\{0,1\}$ the receiver uses any $c\in \{0,1\}$. The players
  perform 1-2 OT with $s_k$, $s_{\bar k}$ and $c$, then the sender
  announces $k$ to the receiver who learns $b$ with probability $1/2$
  if $k=c$. Using their protocol, the receiver can delay his input
  into the box until the sender announced $k$ and always learn $b$.

Buhrman et al. \cite{2006RSPSA.462.1919B}, based on the Wolf and
Wullschleger protocol, showed that bit commitment and OT are possible
given perfect PR-boxes.

The following definition will be of use in the bit-commitment protocol
described in Buhrman et al.

\begin{defn}
  Let the operator $|x|_{11}$ for a bit string $x$ denote the number
  of substrings $11$ of $x$ starting at odd positions (with positions
  starting at 1). $|\cdot|_{11}$ is defined recursively as follows
  \begin{itemize}
  \item $|\epsilon|_{11}=0$ where $\epsilon$ is the empty string;
  \item $|abx|_{11}= |x|_{11}+1$ if $ab=11$, $|x|_{11}$ otherwise.
  \end{itemize}
\end{defn}

Burhman et al.'s protocol for bit commitment consists of repeating $k$
times the following commit/reveal scheme:

\hfill \\ Bit-Commitment($c$)
\begin{description}
  \item[Commit]\hfill
    \begin{itemize}
    \item Alice wants to commit to bit $c$. She constructs $x\in
      \{0,1\}^{2n+1}$ by randomly choosing the first $2n$ bits and
        choosing the last bit such that $|x_1\ldots
        x_{2n}|+x_{2n+1}+c$ is even.
      \item Alice inputs the bits $x_1,\ldots x_{2n+1}$ into the
        $2n+1$ PR-boxes. Let $a_1,\ldots, a_{2n+1}$ be the outputs.
      \item Alice computes $A=\bigoplus_i a_i$ and sends it to Bob.
      \item Bob chooses a random string $y\in_R \{0,1\}^{2n+1}$ and
        inputs bits $y_1,\ldots,y_{2n+1}$ into his end of the $2n+1$
        PR-boxes. Let $b_1,\ldots,b_{2n+1}$ be the outputs.
    \end{itemize}
  \item[Reveal]\hfill
    \begin{itemize}
    \item Alice sends $c$, $x$ and $b_1,\ldots,b_{2n+1}$ to Bob.
    \item Bob checks if $a_i\oplus b_i = x_i\cdot y_i$ for $1\leq
      i\leq 2n+1$ and $|x_1\ldots x_{2n}|_{11}+x_{2n+1}+c$ is even. If
      not, he accuses Alice of cheating.
    \end{itemize}
\end{description}

\begin{thm}
  This protocol is secure against Alice. The best probability with
  which Alice can change her mind is $1/2+ 1/2^{k-1}$.

This protocol is secure against Bob. The best probability with which
Bob can learn $c$ before the reveal stage is $1/2+k/2^{n+1}$.
\end{thm}

\section{Generalized Non-local Boxes}

In this section we study a more general class of non-local boxes,
where we extend the set of inputs, the set of outputs, and the number
of participants. We also present some results on the connections
between types of generalized boxes.

\subsection{Arbitrary Input or Output Size}
Consider boxes with binary inputs, but with outputs taken from
arbitrary finite sets. Let $d_x$ denote the number of inputs and
$d_a$ the number of outputs on Alice's side, similarly $d_y$ denotes
the number of inputs and $d_b$ the number of outputs on Bob's
side. Such boxes correspond to definition \ref{defbox} but where $x\in
\{0,\ldots,d_x-1\}$, $y\in \{0,\ldots,d_y-1\}$,
$a\in\{0,\ldots,d_a-1\}$ and $b\in \{0,\ldots,d_b-1\}$. We will refer
to these as \emph{generalized boxes}.

\subsubsection{$d$-Output Boxes}

The class of generalized boxes with $d_x=d_y=2$ form a polytope
$\mathcal{P}$ described by Barrett et
al. \cite{2005PhRvA..71b2101B}. It's dimension is $4d_ad_b- 2d_a-
2d_b$. So if $d_a=d_b=d$, the dimension is $4d^2-4d$ and when $d=2$ we
find the dimension of the non-signalling polytope of section
\ref{nspoly}.

They also found that the non-local vertices of this polytope are all
equivalent under reversible local operations. A result analogous to
the fact that all non-local vertices of the two-input two-output
polytope are equivalent to the PR-box. 

\begin{thm}
Every non-local vertex of $\mathcal{P}$ is equivalent under
reversible local operations to
\[
P(a,b|x,y)=
\begin{cases}
  1/k &\text{if } (b-a)\equiv xy \mod k\\ 0 &\text{otherwise.}
\end{cases}
\]
for some $k\in \{2,\ldots,\min\{d_a,d_b\}\}$ where $x,y\in
  \{0,1\}$ and $a,b\in \{0,\ldots,k-1\}$.
\end{thm}
Actually, for every $k$, the box described above is a representative of
an equivalence class of non-local vertices.

When $d_a=d_b=k=d$, this box violates the $d$-dimensional
generalization of the CHSH inequality \cite{PhysRevLett.88.040404} up
to it's algebraic maximum. We will refer to such boxes as $d$-output
boxes.

\subsubsection{$d$-Input Boxes}

Jones and Masanes \cite{PhysRevA.72.052312} characterized the set of
generalized boxes for $d_a=d_b=2$ and arbitrary $d_x$ and $d_y$. Every
class of non-local vertices for a given $d_x$ and $d_y$ is represented
by a box parameterized by two integers $g_x\in \{2,\ldots,d_x\}$ and
$g_y\in \{2,\ldots,d_y\}$. The box's behaviour is non-deterministic
for $g_x$ of the inputs and deterministic for $d_x-g_x$ of the inputs
on Alice's side and analogously for Bob.  We send the reader to their
paper for the detailed description of the box.

\subsection{Interconversions of Non-local Correlations}

This section covers the relations existing between different types of
generalized non-local boxes. Theorems \ref{interconvshit} and
\ref{aewoa8gha0w3} are from \cite{2005PhRvA..71b2101B}.

\begin{thm}\label{interconvshit}
  The following interconversions are possible:
  \begin{itemize}
  \item 1 $d$-output box and 1 $d'$-output box can simulate 1
    $dd'$-output box
  \item 1 $dd'$-output box can simulate 1 $d$-output box
  \item $n$ $d$-output boxes can approximate 1 $d'$-output box
  \end{itemize}
\end{thm}

\begin{lem}\label{aewoa8gha0w31}
  Using $n$ $d$-output boxes, Alice and Bob can exactly simulate at
  most $n$ $d'$-output boxes, for $d\geq d'$.
\end{lem}

\begin{lem}\label{aewoa8gha0w32}
  Using $n$ $d'$-output boxes, Alice and Bob can exactly simulate at
  most $n(1+\log_2d')/(1+\log_2d)<n$ $d$-output boxes for $d'\leq d$.
\end{lem}

\begin{thm}\label{aewoa8gha0w3}
  It is in general impossible, using local reversible operations, to
  exactly simulate $m$ $d'$-output boxes from $n$ $d$-output boxes.
\end{thm}

Theorem \ref{aewoa8gha0w3} follows from lemmas \ref{aewoa8gha0w31} and
\ref{aewoa8gha0w32}. It implies that there is little
interconvertibility between families of $d$-output boxes.

The following results by Dupuis et al \cite{2007JMP....48h2107D}
furthers this lack of interconversions by providing impossibilities of
interconversions for $d$-boxes. This first theorem states that any
finite amount of PR-boxes cannot exactly simulate a single 3-box.

\begin{thm}
  It is impossible to simulate a 3-box exactly using a finite
  number of 2-boxes, infinite shared randomness and no communication.
\end{thm}

Their next theorem generalizes their first one. 

\begin{thm}
  Let $S$ be a finite set of generalized non-local box with
  $d_x=d_y=2$ and arbitrary $d_a$ and $d_b$. Then there exists $p$
  such that the $p$-box cannot be simulated by a finite number of
  boxes taken from the set $S$.
\end{thm}

The following results by Jones and Masanes \cite{PhysRevA.72.052312}
show that, to the contrary, $d$-input boxes are very interconvertible.

\begin{thm}
  PR-boxes are sufficient to simulate all non-signalling correlations
  with binary output ($d_a$=$d_b$=2).
\end{thm}

\begin{thm}
  All correlations with arbitrary $d_x$ and $d_y$, and binary output
  ($d_a$= $d_b$=2) are interconvertible.
\end{thm}

We refer the reader to the original paper for both proofs.

\subsection{Multi-party Correlations}

Now consider the case where three participants Alice, Bob and Charlie
exhibit non-local correlations. The definition of the non-local box
can be extended to accommodate this new model.

\begin{defn}
A tripartite correlated box (or box) is a device with three ends. Each
end has the following input-output behaviour: given input $x$, $y$ and
$z$, the box will respectively output $a$, $b$ and $c$ according to
some probability distribution $P(abc|xyz)$ where $x,y,z,a,b,c
\in \{0,1\}$.
\end{defn}

The probabilities $P(abc|xyz)$ are subject to positivity and
normalization, and the trivial extension of the non-signalling
constraints
\[
\sum_a P(abc|xyz)= \sum_a P(abc|x'yz)\quad \forall b,c,x,x',y,z
\]
\[
\sum_b P(abc|xyz)= \sum_b P(abc|xy'z)\quad \forall b,c,x,y,y',z
\]
\[
\sum_c P(abc|xyz)= \sum_c P(abc|xyz')\quad \forall b,c,x,y,z,z'
\]

While the non-signalling condition is roughly unchanged, non-locality
needs to be defined differently than with bipartite
correlations. Alice, Bob and Charlie can be pairwise local which each
other. 
\begin{defn}
   A box $P$ \emph{fully local} if it can be written as
\[
P(abc|xyz)=\sum_i \lambda_i P_i^A(a|x)P_i^B(b|y)P_i^C(c|z)
\]
where $\lambda_i\geq0$ and $\sum_i\lambda_i=1$.
\end{defn}
Can also occur the situation where Alice and Bob are non-local, but
they are local versus Charlie. We call such boxes \emph{two-way local}
along with any box which is a convex combination of such boxes.
\begin{defn}
  A box $P$ is \emph{two-way local} if it can be written as
  \begin{align*}
    P(abc|xyz) &= p_1 \sum_i \lambda_iP_i^{AB}(ab|xy)P_i^C(c|z)
    \\ & + p_2 \sum_i \lambda_iP_i^{AC}(ac|xz)P_i^B(b|y)
    \\ & + p_3 \sum_i \lambda_i P_i^{BC}(bc|yz)P_i^A(a|x)
  \end{align*}
where the $p_i$s and $\lambda_i$s are positive and normalized.
\end{defn}

The set of non-local tripartite correlations form a 26 dimensions
polytope.The set of local correlations form a sub-polytope of the
two-way local polytope, itself a sub-polytope of the non-signalling
polytope.

Vertices of the local polytope correspond to boxes for which all
outputs are deterministic, they are equivalent under reversible local
operations to
\[
P(abc|xyz)=
\begin{cases}
  1 &\text{if } a=0,b=0,c=0 \\
  0 &\text{otherwise.}
\end{cases}
\]
Two-way local vertices are boxes that describe a PR-box shared between
two players while the third has a deterministic box. They are
equivalent under reversible local operations to
\[
P(abc|xyz)=
\begin{cases}
  1/2 &\text{if } a\oplus b=xy\text{ and } c=0\\
  0 &\text{otherwise.}
\end{cases}
\]
Non-local vertices are more complex than the two other types. The set
of non-local vertices has 44 different classes of vertices, which we
won't enumerate. One of these class is equivalent under reversible
local operations to the natural extension of the non-local box
\[
P(abc|xyz)=
\begin{cases}
  1/4 &\text{if } a\oplus b\oplus c= xyz\\
  0 &\text{otherwise.}
\end{cases}
\]

As with the generalized bipartite non-local boxes, it is possible to
perform conversions between tripartite boxes. One could also be
interested in the simulation of tripartite boxes using
PR-boxes.We will, however, not go into further details about these.

\section{Acknowledgements}

Thanks to Gilles Brassard for financial support and to Alain Tapp for
his mentorship.

\bibliography{NLB.bib}

\end{document}